# Studying the internal structures of the central region of prestellar core L1517B in Taurus molecular cloud using ammonia (NH$_3$) (1,1) and (2,2) lines

Atanu Koley[1]

[1] Departamento de Astronomía, Universidad de Concepción, Casilla 160-C, Concepción, Chile
**Author for correspondence:** A. Koley, Email: atanuphysics15@gmail.com.

**Abstract**
Measurement of internal structures in the prestellar core is essential for understanding the initial conditions prior to star formation. In this work, we study the ammonia lines (NH$_3$) ($J$, $K$=1,1 and 2,2) in the central region of the prestellar core L1517B with the Karl G. Jansky Very Large Array Radio Telescope (VLA, spatial resolution $\sim 3.7''$). Our analysis indicates that the central region of the core is close-to-round in shape obtained both from NH$_3$ (1,1) and (2,2) emissions. Radially averaged kinetic temperature ($T_k$) is almost constant with a mean value of $\sim 9$ K. A radially sharp decrease in kinetic temperature ($T_k$) has not been observed inside the central dense nucleus of this prestellar core. In addition, we also notice that there is an overall velocity gradient from north-east to south-west direction in this region, which may be indicative of the rotational motion of the core. We then calculate the parameter β, which is defined as the ratio of rotational energy to gravitational potential energy and find that β equals to $\sim 5 \times 10^{-3}$; which indicates that rotation has no effect at least inside the central region of the core. We also perform the viral analysis and observe that the central region may be in a stage of contraction. From this study, we also show that turbulence inside the central region is subsonic in nature (sonic Mach number, $M_s$ < 1) and has no prominent length-scale dependence. Furthermore, we notice that the decrement of excitation temperature ($T_{ex}$) and column density of NH$_3$ from the centre of the core to the outer side with the peak values of $\sim 5.6$ K and $\sim 10^{15}$ cm$^{-2}$ respectively. In conclusion, this work examines different physical and kinematical properties of the central region of the L1517B prestellar core.

## 1. Introduction

Star formation is one of the fundamental processes in the cosmos. Stars are born in dense cold places called cloud cores. Cores are classified as prestellar or protostellar depending on whether they contain an embedded protostar. Along with gravity, other effects like turbulence, magnetic field, rotation, and temperature form a complex enviornment prior to the star formation (Barranco and Goodman 1998; Kudoh and Basu 2014; Dunham et al. 2016; André 2017; Dobashi et al. 2018; Caselli et al. 2019; Koley and Roy 2019; Dobashi et al. 2019; Tokuda et al. 2020; Sahu et al. 2021; Koley et al. 2021; Koley et al. 2022; Koley 2023). Therefore, studying these quantities are very much essential for a complete understanding of the star formation process. Dust continuum emission provides the line-of-sight averaged temperature of the core (Scott Schnee et al. 2010). Apart from that, ammonia molecule is also used to probe the temperature of the molecular cloud; hence it is often called an "interstellar thermometer". Ammonia molecule exhibits several hyperfine lines. Here, due to the several well-separated lines, measurements of essential parameters like excitation temperature ($T_{ex}$), kinetic temperature ($T_k$), optical depth (τ), column density ($N$), etc., have been possible (Walmsley and Ungerechts 1983; Mangum and Shirley 2015). In addition to these parameters, if one assumes that the non-thermal broadening is caused due to turbulence, then from this spectral study, turbulence measurement is also possible; in particular, about the nature of the turbulence and its scale-dependent behavior. For several decades, temperature measurements have been carried out using ammonia molecules in low and high mass prestellar cores and star-forming regions, both from single-dish and from interferometric telescopes (Barranco and Goodman 1998; Pillai et al. 2006; Dirienzo et al. 2015; Krieger et al. 2017). Although from several studies, it has been argued that in interferometric telescopes a significant amount of flux is resolved out in the prestellar cores (Crapsi et al. 2007; Roy et al. 2011; Koley 2022); for tracing the internal cold dense structures, interferometric observations are generally more effective than single-dish studies (Barranco and Goodman 1998; Crapsi et al. 2007; Dobashi et al. 2018; Tokuda et al. 2020; Sahu et al. 2021).

In this work, we study the internal structures of the central region of L1517B core with the Jansky Very Large Array (VLA) radio telescope. This core was observed previously with several single-dish telescopes using different molecules (Tafalla et al. 2004a; Chitsazzadeh 2014; Megías et al. 2023), where the field of view was larger than our study. Here, we particularly concentrate on the central dense region of the core using the high spatial resolution interferometric telescope.

In section 2, we present the observation and data analysis of our work. In section 3, we mention the ammonia line fitting using *pyspeckit* code (Ginsburg and Mirocha 2011). In section 4, we discuss the main results. At last, in section 5, we summarize our main findings.

## 2. Observation and data analysis

All the observational data were taken from the VLA archival data center. These observations were carried out in 2013 from March 20 to 26 over three observing sessions using the VLA D configuration in the K band (proposal id: VLA-13A/394). Right ascension (R.A.) and declination (Dec.) of the pointing center is $04^h55^m18.00^s$ and $+30°37'46.99''$ respectively (epoch J2000). The primary objective of this study is to analyse the internal structures of the central region of the core using NH$_3$ (*J,K*=1,1 and 2,2) spectral lines. Rest frequencies of these lines are 23.69449550 GHz and 23.722633600 GHz respectively. The bandpass calibrator was 3C48 (0137+331) for this observation, whereas J0414+3418 was used as a phase calibrator. Channel resolution was $\sim$ 3.906 kHz ($\sim$ 0.049 km sec$^{-1}$) with a total number of channels was 2048. For every 4 minutes 24 seconds of source observation, phase calibrator, J0414+3418 was observed for almost 1 minute 45 seconds. In each observing session, bandpass calibrator, 3C48 was observed for $\sim$ 9 minutes 54 seconds. After observing almost 5 hours and 18 minutes on the source, the achieved RMS noise was $\sim$ 2 mJy beam$^{-1}$. This value of RMS noise represents the RMS noise per channel. Initial flagging and calibration were done using the scripted Common Astronomy Software Applications (CASA) pipeline (scripted pipeline version 1.4.2) of the National Radio Astronomy Observatory (NRAO)[a]. Additional flagging was needed due to the bad amplitude gain of a few antennas. We would like to emphasize that we have not done the hanning smoothing of the UV data while calibrating using the scripted pipeline. With the calibrated data produced by the pipeline, we performed the cleaning with the CASA task **TCLEAN** (CASA version 5.3.0). Please note that, we ran it separately for the NH$_3$ (*J,K*=1,1) and NH$_3$ (*J,K*=2,2) lines. Here, we used the deconvolver "multiscale" with the scales 3.2″, 6.4″, 12.8″, 25.6″, and 51.2″ respectively to pick up all the structures. We used the robust parameter, **ROBUST** = +2 for natural weighting. The synthesized beam that is convolved with the clean components is $\sim 4.01'' \times 3.37''$ with a position angle of + 52.02°. Final image cubes (both for 1,1 and 2,2 lines) produced after cleaning are used to study the core using the simultaneous fitting of these lines. Here we would like to point out that due to the use of interferometric-only observations, our data are lacking a contribution from extended NH$_3$ emissions. This effect may affect the determination of parameters that depend on the absolute intensity of the lines, such as the column density of NH$_3$. However, it may be less severe for parameters that are dependent on the line ratio, such as the kinetic temperature ($T_k$) of the gas.

---

a. The National Radio Astronomy Observatory is a facility of the National Science Foundation operated under cooperative agreement by Associated Universities, Inc.

## 3. Line fitting

For fitting the NH$_3$ (*J,K*=1,1) and NH$_3$ (*J,K*=2,2) lines and in turn obtaining different useful informations mentioned earlier, we use the *pyspeckit* code (Ginsburg and Mirocha 2011). This code is based on the nonlinear gradient descent algorithm (MP-FIT, Markwardt 2009), where various useful parameters are obtained after comparing the simulated and observed brightness temperature ($T_B$) of the lines. For a detailed discussion regarding the fitting, we refer the study of Friesen et al. (2017). Initial guesses in terms of kinetic ($T_k$) or rotational temperature ($T_{rot}$), excitation temperature ($T_{ex}$), para column density ($N_{para}$), ortho ratio ($F_{ortho}$), centre velocity ($v_c$), and the velocity dispersion ($\sigma_{total}$) of the line have to be put in the fitting. As the line modeling crucially depends on the centre velocity ($v_c$), its initial guess has been calculated from the moment map of the cube before the fitting. We note that the final results do not alter as long as the initial parameters are within a reliable range. We would also like to point out that we took only those pixels where the signal-to-noise ratio (SNR) of the NH$_3$(1,1) emission is greater than 5.5. In the modeling, it is assumed that $T_{ex}$ is the same for both (1,1) and (2,2) lines. Out of different types of fitters, we use the *fittype* = 'cold_ammonia', which makes the assumption that for typically low temperature in the prestellar cores, only (1,1), (2,2) and (2,1) lines of para-NH$_3$ are occupied (Krieger et al. 2017; Pineda et al. 2021). It is also assumed that the ortho-to-para ratio of ammonia is 1. This code fits pixel-wise spectra and finally forms a cube, where various fitted parameters and their associated errors are stored in different planes, which are further used for detailed analyzing the core. For a detailed understanding of the different energy levels of NH$_3$ lines, we refer the works of Walmsley and Ungerechts (1983) and Mangum and Shirley (2015).

## 4. Results

### *4.1 Integrated intensity emission & column density*

Left panel of Fig. 1 shows the overplot of integrated intensity map of NH$_3$ (1,1) emission both in color and contour plots. Likewise, the right panel of Fig. 1 shows the overplot of integrated intensity map of NH$_3$ (2,2) emission both in color and contour plots. Based on these figures, it is apparent that NH$_3$ (2,2) emission is less spread than NH$_3$ (1,1) emission. This is supported by the previous single-dish observation of this core (Chitsazzadeh 2014). From these figures, it is also noticeable that NH$_3$ (1,1) and (2,2) emissions are centrally concentrated, but the peak positions of the integrated intensity maps have a slight offset compared to the dust and H$_2$ column density maps where the peaks of them are identical (Tafalla et al. 2004a; Megías et al. 2023). The location of the dust peak is $04^h55^m17.60^s$, $+30°37'44.00''$, whereas the peaks of the NH$_3$ (1,1) and (2,2) integrated intensity maps are $04^h55^m18.31^s$, $+30°37'42.38''$ and $04^h55^m18.06^s$, $+30°37'41.34''$ respectively. It is interesting to note that, the central region of the core is close-to-round in shape, rather than a filament-like structure (Kauffmann et al. 2008; Dobashi et al. 2018). Since the shape of the central region of the core is round, an average of various essential parameters can be calculated over concentric circles to obtain the radially averaged profiles of the essential parameters. On the other hand, if the structure had elongated, an average would have to be performed over successive ellipses (with a fixed aspect ratio and position angle). From the line fitting using *pyspeckit* code, we obtain the column density map of ammonia emission in the L1517B core, which is shown in the left panel of Fig. 2. Using this distribution, we obtain the radially averaged profile of column density in the core. Here, we would also like to point out that, we consider the peak position of the dust continuum emission as the centre of the core. This is because the peak of dust and the peak of $H_2$ column density match well in this core (Megías et al. 2023). And for converting the angular scale into the physical distance on the plane of the sky, we take 159 pc distance of the L1517B core (Galli et al. 2019). Right panel of Fig. 2 represents the radially averaged column density profile. From this figure, it indicates that the peak column density of $NH_3$ is $\sim 10^{15}$ cm$^{-2}$ and decreases towards the outer edge. We also note that the distance where the column density decreases by a factor of 2 of its peak value is at radius $\sim 0.016$ pc.

### 4.2 Spectral properties across the core

In Fig. 3, we show the spectra of $NH_3$ (1,1) and (2,2) lines towards different positions inside the core. For example, we show the spectra towards the dust peak position. Similarly, we also plot the spectra towards the peaks of $NH_3$ (1,1) and (2,2) emissions. Here, it is interesting to note that the blue-skewed profile that is observed in the (1,1) spectra is not due to the collapse (Myers et al. 1996) or any other cloud dynamics (Evans 1999), rather it is the intrinsic closely separated lines caused by the hyperfine splitting. Here, we would like to point out that all the calibrated data and the *pyspeckit* results are obtained from the original data. No hanning smoothing has been performed in the entire analysis except for these spectra in Fig. 3. As a simple matter of making the spectra more visible, we use the CASA viewer task to smooth the spectra in Fig. 3.

Now, we discuss the Left panel of Fig. 4, where we show the velocity field across the region. From this figure, it appears that there is an indication of overall velocity gradient from north-east to south-west direction. This might be caused due to the rotational motion of the core. However, the exact pattern is more complex than a continuous increase or decrease in velocity of equal magnitude. Velocity gradient across the core was noticed in the earlier single-dish observations with the $NH_3$ (1,1) and $N_2H^+$ (1-0) lines where the fields of view were larger than our study (Goodman et al. 1993; Tafalla et al. 2004a; Chitsazzadeh 2014). Right panel of Fig. 4 shows the overplot of the local velocity gradient and velocity field across the region. Here, in the color plot we subtract the systematic velocity of the core (+ 5.79 km sec$^{-1}$) from the velocity field. Now, in order to measure the overall velocity gradient, we first calculate the velocity gradient at each position and finally obtain the overall gradient in this region. Detailed discussion regarding the calculation of local and overall velocity gradient is mentioned in Appendix 1. From the analysis, we obtain the overall velocity gradient, $\psi \sim 1.10$ km sec$^{-1}$ pc$^{-1}$ and direction, $\theta_{\hat{\psi}} \sim 127°$ west of north. This value is similar to the earlier single-dish observations studied by Goodman et al. 1993 and Tafalla et al. 2004a, where the measured values were 1.52 and 1.10 km sec$^{-1}$ pc$^{-1}$ respectively. However, this value is one order smaller than the L1544 core, where the value is $\sim 9.0$ km sec$^{-1}$pc$^{-1}$ based on the interferometric (VLA) observations of $NH_3$ (1,1) and (2,2) lines (Crapsi et al. 2007).

If there is a strict continuous velocity gradient of equal magnitude across the region, it is possible to conclude strongly that the region is rotating. However, in our region, both visually and in terms of the magnitude of the overall velocity gradient, it appears to be a rotation, despite not meeting the strict criteria of smooth and continuous velocity variation. It is also true that there are many challenges associated with the analysis and fitting of real astronomical data. Consequently, obtaining a smooth variation of velocity across the region is extremely difficult. Therefore, in the following, we calculate the ratio of rotational energy to gravitational potential energy, assuming that the velocity gradient is the cause of rotation. This enables us to gain a rough understanding of the role of rotation in this region if the overall velocity gradient is caused by the rotation. The parameter β, which is the ratio of the rotational energy to the gravitational energy (Goodman et al. 1993) is defined by the formula:

$$\beta = \frac{\frac{1}{2} I\omega^2}{q \frac{GM^2}{R}} = \frac{1}{2} \frac{p}{q} \frac{\omega^2 R^3}{GM} \quad (1)$$

Here, $I$ is the moment of inertia, $\omega$ is the angular velocity, $G$ is the gravitational constant, $M$ is the mass, $R$ is the radius of the core, $p$ and $q$ are unit-less numbers, which vary depending on the geometry and the density profile of the system (Kauffmann, Pillai, and Goldsmith 2013). The value of $(\frac{p}{q})$ is 0.66 for a constant density sphere. However, when the density profile varies as $r^{-2}$ with fixed $M$ and $R$, this value is 0.22, which is one third of the earlier one (see Appendix 2). From the fitting of observed continuum emission, Tafalla et al. (2004a) showed that the number density of this core is not constant rather follows a power law: $n(r) = \frac{n_0}{1+(\frac{r}{r_0})^{2.5}}$. Here, $r$ is the radial distance, $n_0 = 2.2 \times 10^5$ cm$^{-3}$ and $r_0 = 35.00''$ or 0.027 pc. Since we are only analyzing the central 0.025 pc region inside the core, we consider $\frac{p}{q} = 0.66$ for our analysis, which will not lead to a significant difference in the result. Likewise, we consider only the enclosed mass within the radius of 0.025 pc. From the work of Benson and Myers 1989, the mass of the core is $\sim 0.50$ M$_\odot$, where it was assumed a radius of $\sim 0.068$ pc (assuming 159 pc distance) based on the observation of $NH_3$ (1,1) line. Now, according to the power-law of density ($r^{-2.5}$), the mass enclosed in a radius of 0.025 pc is $\sim 0.30$ M$_\odot$. Furthermore, the value of $\omega$ is obtained from $\psi$ after considering the inclination angle ($i$) of $\omega$ to the line-of-sight. However, for a single measurement, we have not taken into account this statistical correction. This correction factor is small and will

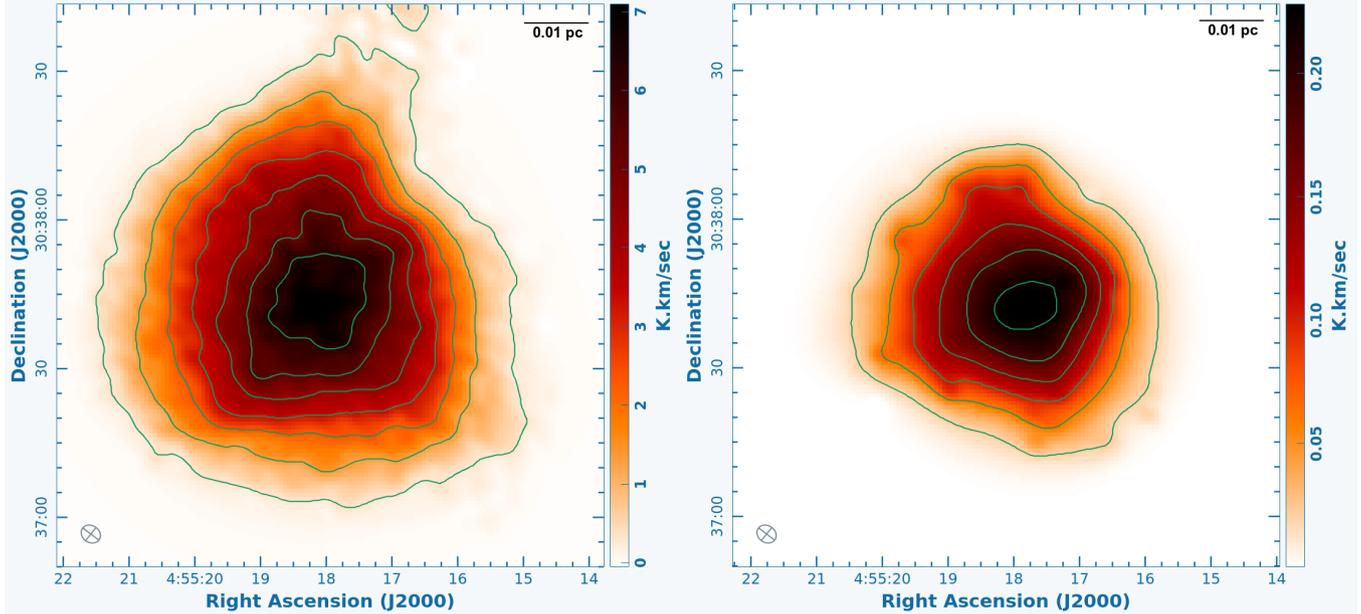

**Figure 1.** Left: Integrated intensity map of $NH_3$ (1,1) emission both in color and contour plots (integrated over main and satellite lines). The contour levels are 0.30, 1.30, 2.30, 3.30, 4.30, 5.30, and 6.30 K.km sec$^{-1}$ respectively. Right: Integrated intensity map of $NH_3$ (2,2) emission both in color and contour plots. The contour levels are 0.02, 0.06, 0.10, 0.14, 0.18, and 0.22 K.km sec$^{-1}$ respectively.

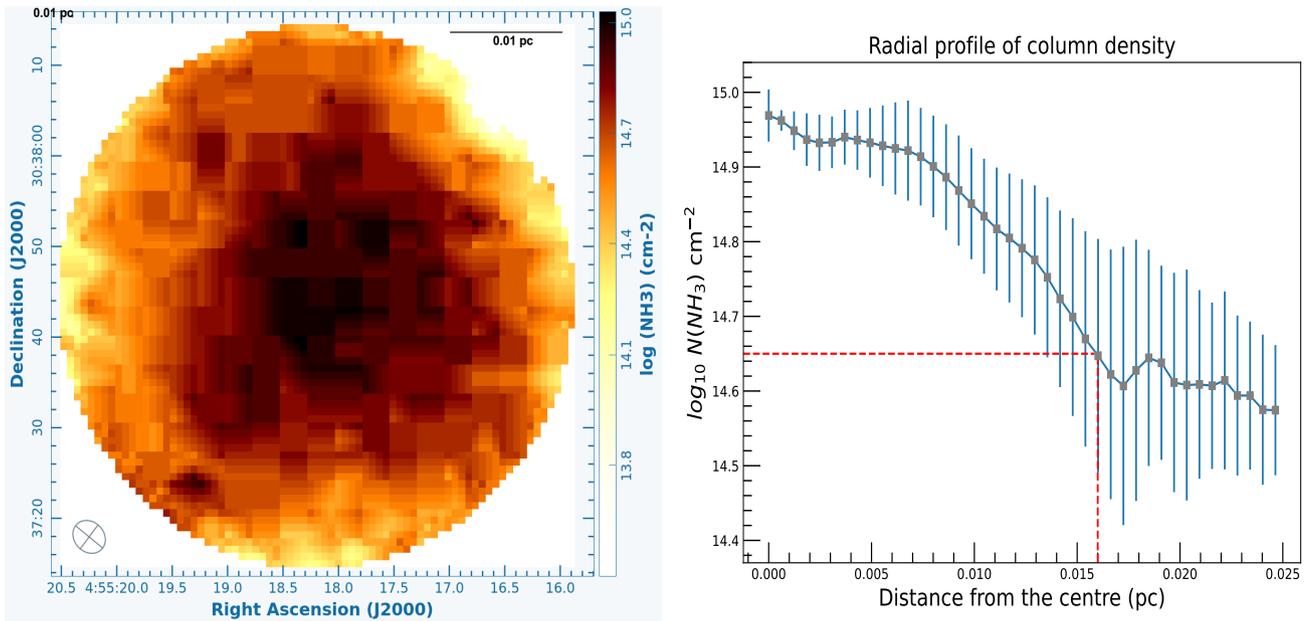

**Figure 2.** Left: Column density ($\log_{10} N(NH_3)$) distribution of the L1517B core. Right: Radial profile of $\log_{10} N(NH_3)$ inside the core. Grey squares are the mean values for successive concentric circles around the centre, whereas the vertical blue lines on both sides are the $1\sigma$ uncertainties of the mean values. Two dashed red lines intersect the curve at the point where column density drops by a factor of two, which is at 0.016 pc.

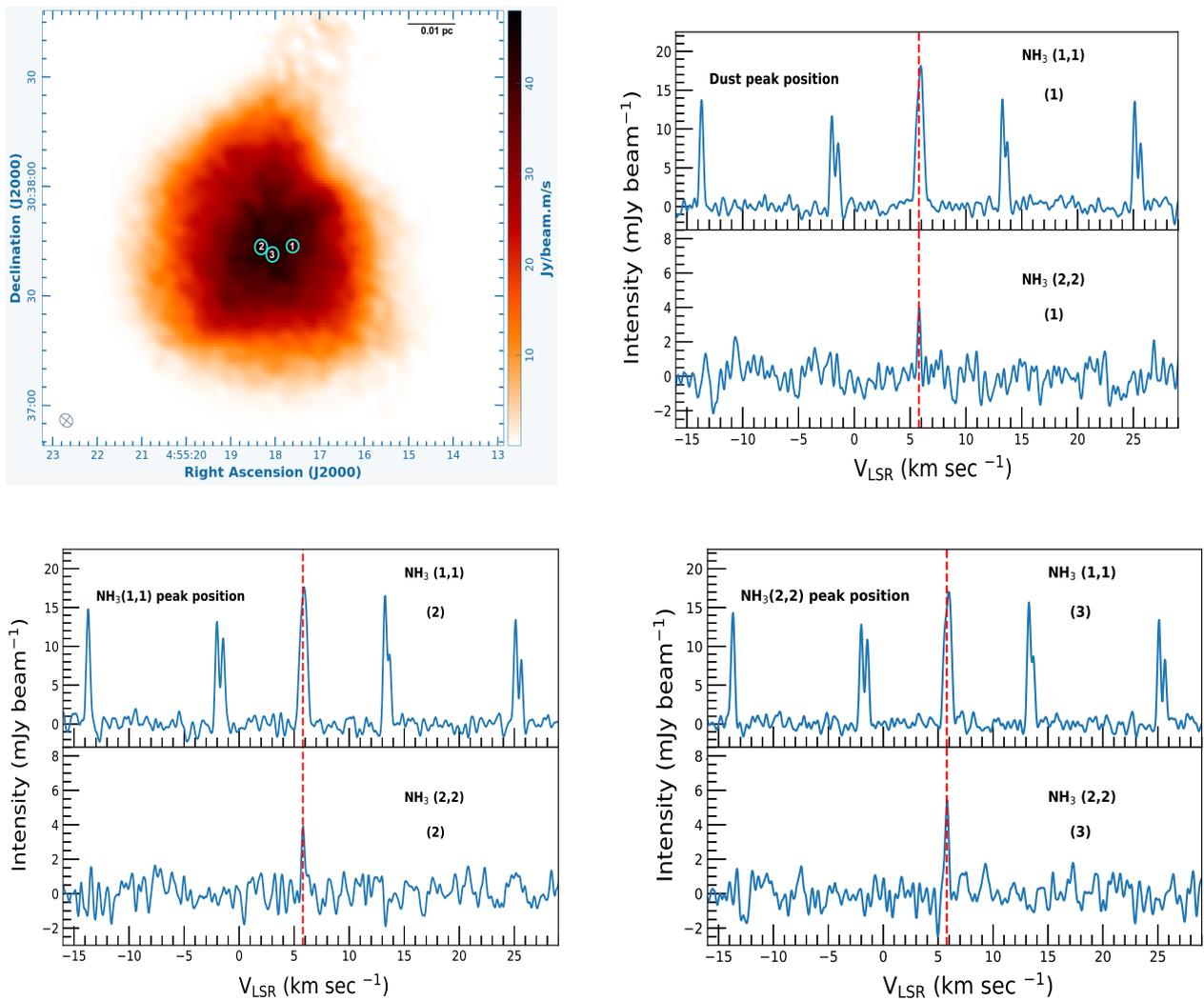

**Figure 3.** Upper panel: Left: Integrated intensity map of NH$_3$ (1,1) emission in color plot. Here, different circles marked with numerical numbers denote the positions where the spectra of (1,1) and (2,2) lines are shown in the following figures. Right: Spectra of NH$_3$ (1,1) and (2,2) lines (after hanning smoothing with kernel width 11) towards the dust peak position ($04^h55^m17.60^s$, $+30°37'44.00''$) marked by **1** in the left figure. Here, the red dashed vertical line corresponds to the systematic velocity of the core, which is + 5.79 km sec$^{-1}$. Lower panel: Left: Same as uper panel right figure but towards the NH$_3$ (1,1) peak position ($04^h55^m18.31^s$, $+30°37'42.38''$) marked by **2** in the uper panel left figure. Right: Same as left but towards the NH$_3$ (2,2) peak position ($04^h55^m18.06^s$, $+30°37'41.34''$) marked by **3** in the uper panel left figure.

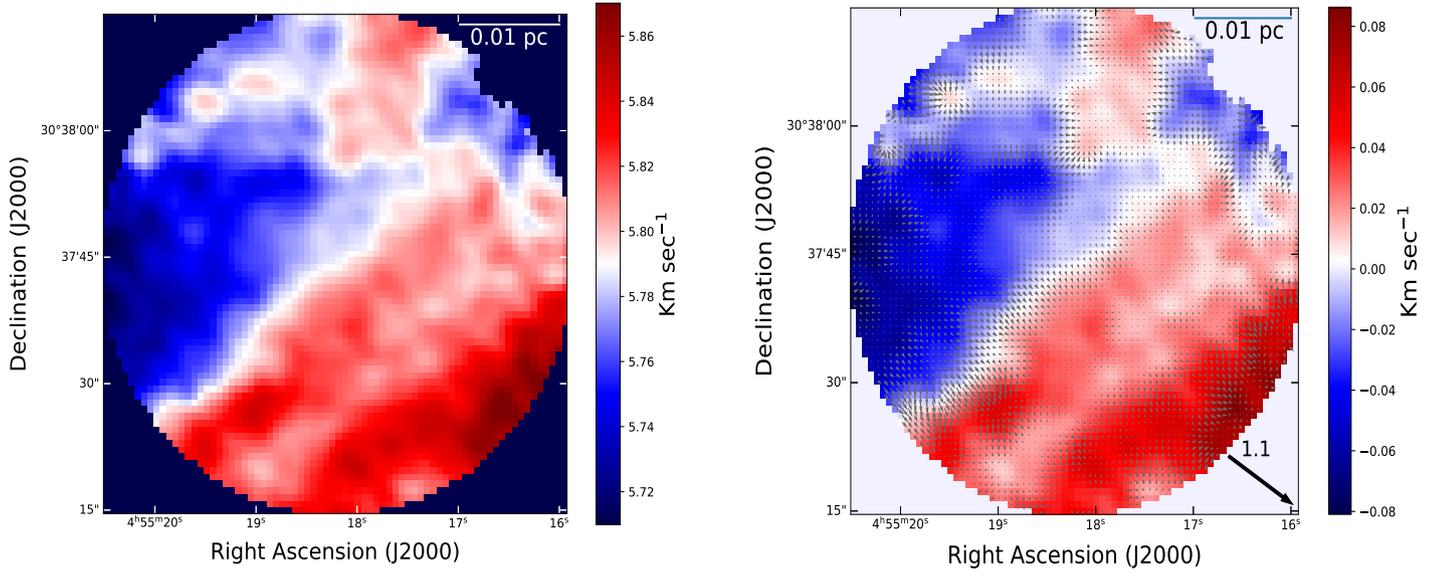

**Figure 4.** Left: Centre velocity ($V_c$) distribution of the L1517B core. Right: Overplot of velocity fields and velocity gradient vectors (10″ corresponds to a gradient of 89.28 km sec$^{-1}$ pc$^{-1}$) across the region. Here, the systematic velocity of the core (+5.79 km sec$^{-1}$) has been subtracted from the centre velocity. The black solid arrow on the right bottom corner represents the overall velocity gradient ($\hat{\psi} \sim 1.10$ km sec$^{-1}$ pc$^{-1}$, 10″ corresponds to 1.0 km sec$^{-1}$ pc$^{-1}$) and its direction ($\theta_{\hat{\psi}} \sim 127°$ west of north).

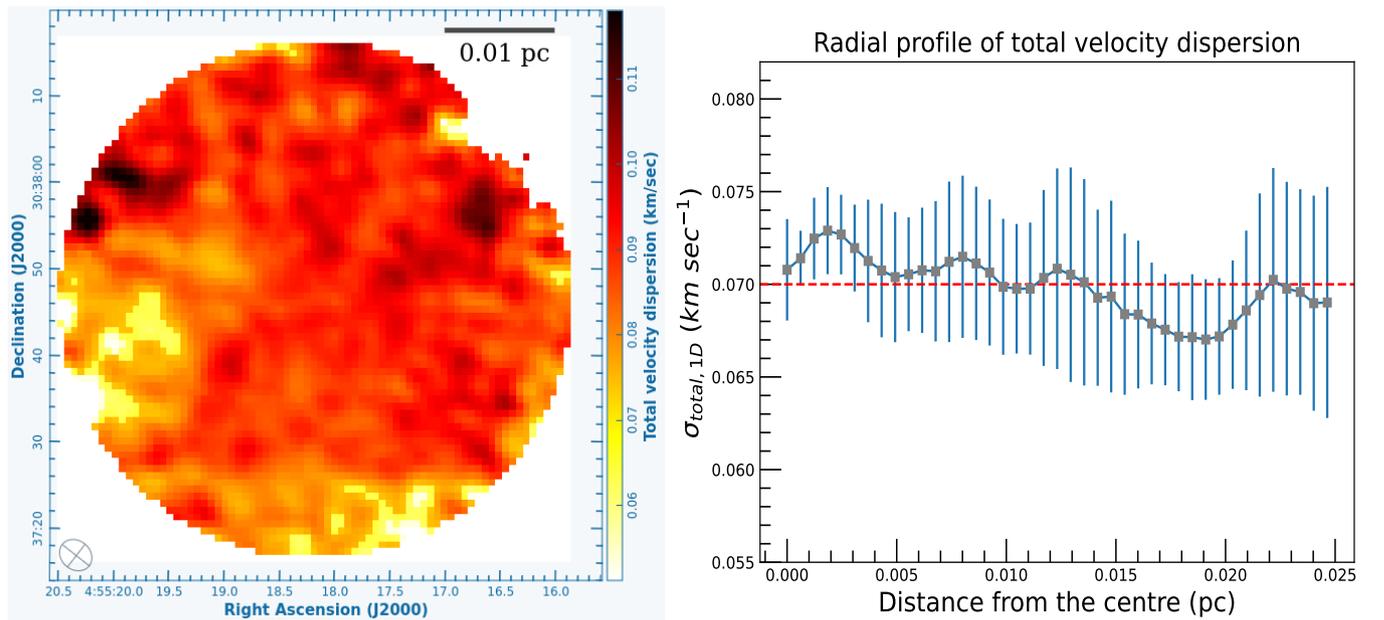

**Figure 5.** Left: Distribution of total velocity dispersion ($\sigma_{\text{total,1D}}$) of the L1517B core. Right: Radial profile of $\sigma_{\text{total,1D}}$ inside the core. Grey squares are the mean values for successive concentric circles around the centre, whereas the vertical blue lines on both sides are the 1$\sigma$ uncertainties of the mean values. Here, the red dashed horizontal line represents the line at which $\sigma_{\text{total,1D}}$ is 0.07 km sec$^{-1}$.

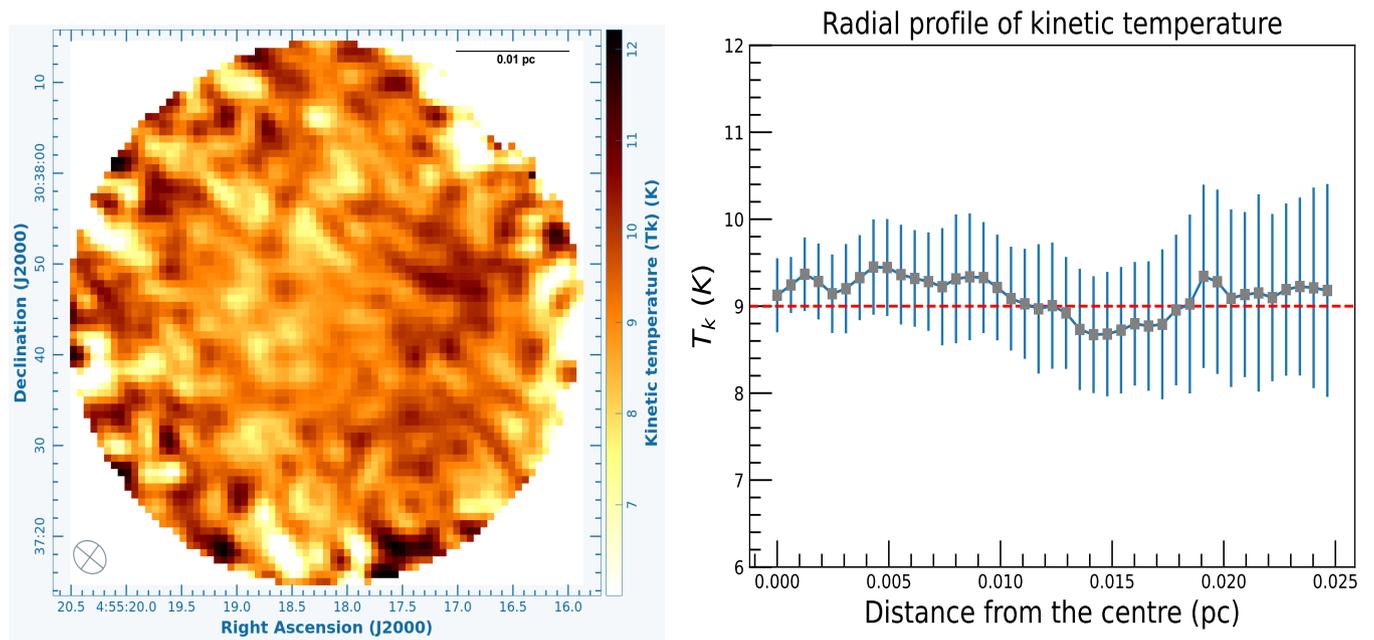

**Figure 6.** Left: Kinetic temperature ($T_k$) distribution of the L1517B core. Right: Radial profile of kinetic temperature ($T_k$) inside the core. Grey squares are the mean values for successive concentric circles around the centre, whereas the vertical blue lines on both sides are the 1$\sigma$ uncertainties of the mean values. Here, the red dashed horizontal line represents the line at which $T_k$ is 9.0 K

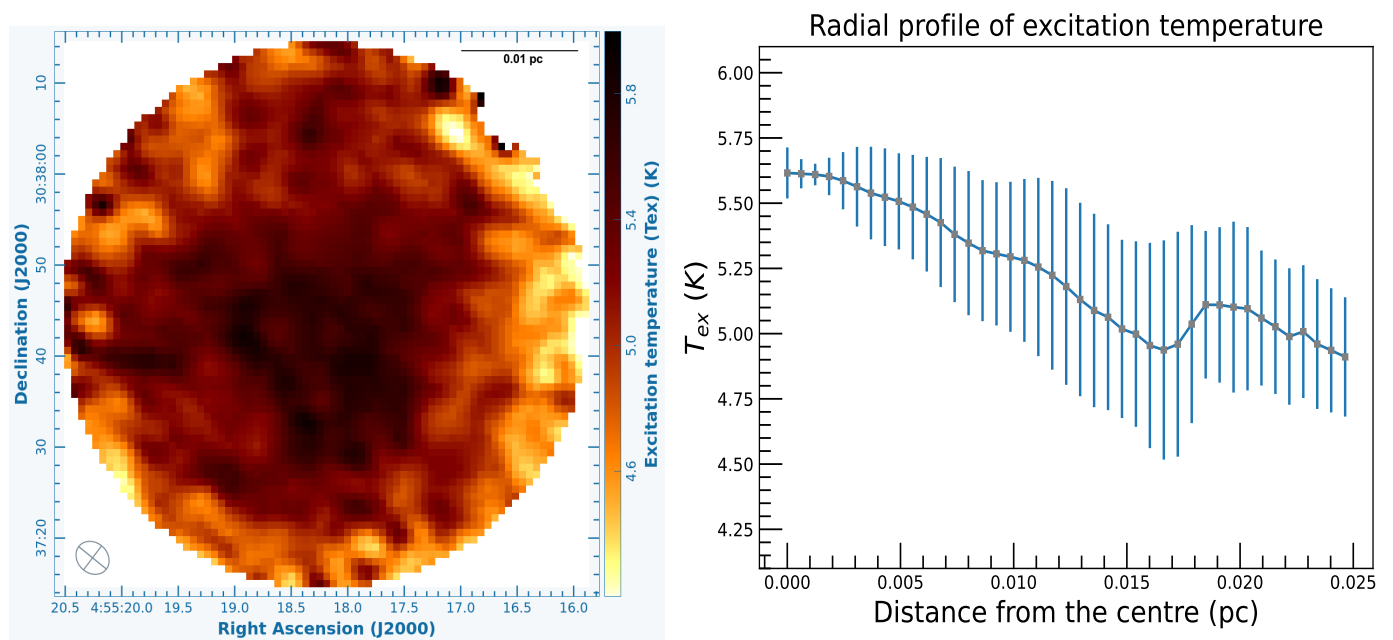

**Figure 7.** Left: Excitation temperature ($T_{ex}$) distribution of the L1517B core. Right: Radial profile of $T_{ex}$ inside the core. Grey squares are the mean values for successive concentric circles around the centre, whereas the vertical blue lines on both sides are the 1$\sigma$ uncertainties of the mean values.

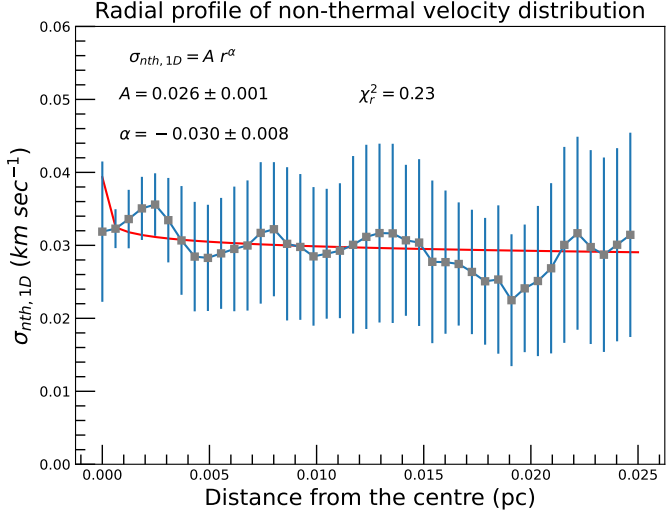

**Figure 8.** Radial profile of non-thermal velocity dispersion ($\sigma_{nth,1D}$) inside the core. Grey squares are the mean values for successive concentric circles around the centre, whereas the vertical blue lines on both sides are the 1$\sigma$ uncertainties of the mean values. Here, red solid line is the best fitted $\sigma_{nth,1D}$ curve.

not significantly affect the value of β (Goodman et al. 1993). After putting all these in equation 1, we finally obtain β equals to $\sim 5 \times 10^{-3}$. From this value, it indicates that rotation has no effect on the gas dynamics, at least inside the central region of the core.

After that, we show the spatial distribution and the radially averaged profile of total velocity dispersion ($\sigma_{total,1D}$) in the left and right panels of Fig. 5. Here, we would like to point out that we have corrected the $\sigma_{total,1D}$ due to the channel resolution effect, which means $\sigma_{total,1D} = \sqrt{\sigma_{total,1D\,fit}^2 - (\frac{\Delta ch}{2.35})^2}$. Here, $\sigma_{total,1D\,fit}$ is the resulting velocity dispersion from the fit and $\Delta$ch is the channel resolution of the observation (Pineda et al. 2021). From the right figure, it indicates that there is a slight variation of $\sigma_{total,1D}$ with a mean value of $\sim 0.07$ km sec$^{-1}$. We note that this value is lower than that reported earlier from the single-dish study with the same NH$_3$ line emissions, where $\sigma_{total,1D}$ was $\sim 0.09$ km sec$^{-1}$ (Tafalla et al. 2004a, 2004b). Now, it is important to check whether this total (thermal and non-thermal) motion and the magnetic field (B) are capable of preventing the collapse. For that we calculate a parameter W, which is defined by the formula:

$$W = \frac{2 \in_k + \Lambda_B}{\Phi_\nu} \quad (2)$$

Here, $\in_k$ is total kinetic energy of the system, which is $\frac{3}{2} M \sigma_{total,1D}^2$, $\Lambda_B$ is the magnetic energy of the system, which is $\frac{B^2}{6R^3}$, and $\Phi_\nu$ is the potential energy of the system which is equal to $-q \frac{GM^2}{R}$. A value of W equal to 1 indicates a state of virial equilibrium, and a value less than 1 indicates possible gravitational contraction. Kirk, Ward-Thompson, and Crutcher (2006) estimated the magnetic field (B) in L1517B prestellar core using the Chandrasekhar & Fermi (CF) method and obtained $\sim 30 \pm 10$ µG. We take this value of B in our analysis. We also consider the M and R values in the same manner as described above. After putting all these values we finally obtain W equals to 0.55. We note that, due to the uncertainties of M, B, $\sigma_{total,1D}$, and R, this parameter W will vary. For example, if we take B = 20 µG instead of 30 µG, we obtain W equals to 0.41. Similarly, if we take B = 40 µG, the value of W becomes 0.75. Thus, we can say that this central region may be in a stage of contraction. This is also supported by the previous work of Fu, Gao, and Lou (2011).

### 4.3 Temperature

In the left and right panels of Fig. 6, we show the pixel-wise kinetic temperature ($T_k$) distribution and the radially averaged profile of $T_k$ in the core. From the right figure, it indicates that radially averaged $T_k$ values vary slightly with a mean value of $\sim 9$ K. Therefore, we can argue that there is no such prominent radial temperature gradient exist in the central region of the core. We would like to pint out that, in other highly compact cores (TMC-1C and L1544) (Schnee and Goodman 2005; Crapsi et al. 2007; Sipilä et al. 2022), where the central density is $\sim 10^{6.5-7.0}$ cm$^{-3}$, it has been observed using dust SED (spectral energy distribution) fitting and using the ammonia line that there is a sharp decrease in central kinetic temperature up to $\sim 6$ K. However, in some other prestellar cores, where the central density is $\sim 10^{5.0-5.5}$ cm$^{-3}$, similar to this prestellar core, earlier studies did not find any such sharp decrease in kinetic temperature inside the central region of the prestellar core (Ruoskanen et al. 2011; Chitsazzadeh 2014; Pineda et al. 2022). It is possible that the surrounding environment (in terms of the radiation field) of these highly compact prestellar cores differs from that of the other prestellar cores, which do not experience a sharp drop in kinetic temperature.

In Fig. 7, we show the spatial distribution and radially averaged profile of excitation temperature ($T_{ex}$). From these figures, it shows that $T_{ex}$ increases from the outside to the inside the central nucleus of the core with a peak value of $\sim 5.6$ K. It indicates that due to the increase of the density towards the inner core or nucleus, $T_{ex}$ is more strongly coupled with $T_k$.

### 4.4 Turbulence

Turbulence can be studied from the spectral analysis if one assumes that the non-thermal broadening is caused due to turbulence. If this is the case, then it is obtained by the formula:

$$\sigma_{nth,1D} = \sqrt{\sigma_{total,1D}^2 - \sigma_{th,1D}^2} \quad (3)$$

Here, $\sigma_{total,1D}$ is the total velocity dispersion of the observed line, whereas $\sigma_{th,1D}$ and $\sigma_{nth,1D}$ are the thermal and non-thermal broadenings of the line. Distribution of $\sigma_{total,1D}$ is obtained from the output of *pyspeckit* code whereas, pixel-wise

$\sigma_{\text{th,1D}}$ can be calculated from the measured $T_k$ by the formula (Sepúlveda et al. 2020):

$$\sigma_{\text{th,1D}} = \sqrt{\frac{k_B T_k}{\mu_{NH_3} m_H}} \quad (4)$$

Here, $k_B$ is the Boltzmann constant, $m_H$ is the mass of the hydrogen atom, $T_k$ is the kinetic temperature and $\mu_{NH_3}$ = 17 is the molecular mass of the $NH_3$ molecule.

In Fig. 8, we show the radially averaged $\sigma_{\text{nth,1D}}$ profile. From this figure, it appears that the radially averaged values vary slightly with a mean value of 0.03 km sec$^{-1}$. To check whether the $\sigma_{\text{nth,1D}}$ has any inherent length-scale dependency, we fit the profile with $Ar^\alpha$, where $A$ is the magnitude in km sec$^{-1}$, $r$ is the length-scale in pc and $\alpha$ is the power-law index. The values of $A$ and $\alpha$ come out to be 0.026 ± 0.001 km sec$^{-1}$ and - 0.030 ± 0.008 respectively. These values indicate that there is no prominent length-scale dependence on turbulence. Now, if we consider that the turbulence is isotropic, then total three dimensional velocity dispersion $\sigma_{\text{nth,3D}}$ can be obtained from $\sigma_{\text{nth,1D}}$ after multiplying $\sqrt{3}$. On the other hand, sound speed $c_s$ can also be obtained by:

$$c_s = \sqrt{\frac{k_B T_k}{\mu_{\text{gas}} m_H}} \quad (5)$$

Here, $\mu_{\text{gas}}$ = 2.37 is the average mass of molecular gas. We then finally obtain the sonic Mach number $M_s$ by the formula:

$$M_s = \left(\frac{\sigma_{\text{nth,3D}}}{c_s}\right) \quad (6)$$

From this analysis, radially averaged mean value of $T_k$ is ∼ 9 K, and radially average mean value of $\sigma_{\text{nth,1D}}$ is ∼ 0.03 km sec$^{-1}$. Therefore, $M_s$ ∼ 0.3 inside this central region of the core. From the value of $M_s$, it indicates that turbulence is subsonic in nature. This result supports previous observational studies as well as the theoretical ideas of the dissipation of turbulence near the ambipolar diffusion scale, which is close to the size of the prestellar core (∼ 0.1 pc) (Barranco and Goodman 1998; Stahler and Palla 2004; S. Schnee et al. 2007; Koley 2022). However, a recent observational study reveals interesting results, where they argue that although the turbulence (or perturbed MHD Alfven wave) is dissipated in neutral near the outer edge of the core, ions still carry the turbulence inside the core, and in turn, they have wider line width compared to the neutrals (Pineda et al. 2021). In the future, more observational studies including $N_2H^+$, will ascertain the detailed nature of the turbulence inside the prestellar core.

## 5. conclusions

In this work, we analysis the central region of the prestellar core L1517B with $NH_3$ (1,1) and (2,2) lines with the Jansky Very Large Array (VLA) telescope. Our observations and analysis reveal the following:

(1) The central region of the core is close-to-round in shape obtained both from $NH_3$ (1,1) and (2,2) observations.

(2) Column density of $NH_3$ decreases from the centre of the core to the edge, with a peak value of ∼ $10^{15}$ cm$^{-2}$ and the distance where it decreases by a factor 2 is at radius 0.016 pc.

(3) There is an overall velocity gradient ($\psi$ ∼ 1.10 km sec$^{-1}$ pc$^{-1}$) from the north-east to south-west direction ($\theta_{\hat{\psi}}$ ∼ 127° west of north) in this region, which may cause due to the rotation of the core. This value of $\psi$ is similar to the earlier single-dish observation obtained from $NH_3$ emission. After assuming the velocity gradient caused due to rotation, we calculate the parameter $\beta$, which is defined by the ratio of rotational energy to the gravitational potential energy and obtain $\beta$ equals to ∼ $5 \times 10^{-3}$, which indicates that rotation has no significant effect on gas dynamics, at least inside the central region of the core.

(4) We find that the total velocity dispersion ($\sigma_{\text{total,1D}}$) varies slightly with mean value of 0.07 km sec$^{-1}$. It is slightly smaller than the previously measured value of ∼ 0.09 km sec$^{-1}$ from single-dish observation.

(5) We perform the virial analysis including the magnetic field taken from the previous study and conclude that the central region could be in a stage of collapse.

(6) Kinetic temperature ($T_k$) of this central region is nearly constant with a mean value of ∼ 9 K. No sharp decrease of the $T_k$ has been observed inside the nucleus of the prestellar core.

(7) We observe that the excitation temperature ($T_{ex}$) increases towards the nucleus of the core with a peak value of ∼ 5.6 K. It suggests that due to the increase in the density towards the inner core or nucleus, $T_{ex}$ is more strongly coupled with $T_k$.

(8) We study the turbulence in this core and find that turbulence is subsonic in nature (sonic Mach number, $M_s$ < 1) and has no prominent length-scale dependence. This is supported by earlier studies in several prestellar cores.

**Acknowledgements**
We thank the reviewer for insightful comments and valuable suggestions that have helped in improving this paper significantly. AK gratefully acknowledges support from ANID BASAL project FB210003. The author would like to acknowledge the Indian Institute of Science, Bangalore, India, where most of the research was conducted while AK was working there as a Research Associate (RA). AK would also like to extend his deepest gratitude to Prof. Nirupam Roy for introducing him to the study of ammonia during his earlier research.

*Telescope:* Jansky Very Large Array Telescope (VLA).

*Software:* Aplpy (Robitaille and Bressert 2012), Astropy (Astropy Collaboration et al. 2013), CARTA (Comrie et al. 2021), CASA (Common Astronomy Software Applications package- National Radio Astronomical Observatory), Matplotlib (Hunter 2007), PySpeckit (Ginsburg and Mirocha 2011), Scousepy (Henshaw et al. 2016; Henshaw et al. 2019).

## Data Availability

All data used in this study are available publicly from the VLA online archive. Authors will be happy to share all the reduced data and the velocity gradient code on reasonable request.

## References

André, Philippe. 2017. Interstellar filaments and star formation. *Comptes Rendus Geoscience* 349, no. 5 (September): 187–197. https://doi.org/10.1016/j.crte.2017.07.002. arXiv: 1710.01030 [astro-ph.GA].

Astropy Collaboration, Thomas P. Robitaille, Erik J. Tollerud, Perry Greenfield, Michael Droettboom, Erik Bray, Tom Aldcroft, et al. 2013. Astropy: A community Python package for astronomy. 558 (October): A33. https://doi.org/10.1051/0004-6361/201322068. arXiv: 1307.6212 [astro-ph.IM].

Barranco, Joseph A., and Alyssa A. Goodman. 1998. Coherent Dense Cores. I. $NH_3$ Observations. 504, no. 1 (September): 207–222. https://doi.org/10.1086/306044.

Benson, P. J., and P. C. Myers. 1989. A Survey for Dense Cores in Dark Clouds. 71 (September): 89. https://doi.org/10.1086/191365.

Caselli, Paola, Jaime E. Pineda, Bo Zhao, Malcolm C. Walmsley, Eric Keto, Mario Tafalla, Ana Chacón-Tanarro, et al. 2019. The Central 1000 au of a Pre-stellar Core Revealed with ALMA. I. 1.3 mm Continuum Observations. 874, no. 1 (March): 89. https://doi.org/10.3847/1538-4357/ab0700. arXiv: 1902.05299 [astro-ph.SR].

Chitsazzadeh, Shadi. 2014. Internal Physical and Chemical Characteristics of Starless Cores on the Brink of Gravitational Collapse. PhD diss., University of Victoria, Canada, August.

Comrie, Angus, Kuo-Song Wang, Shou-Chieh Hsu, Anthony Moraghan, Pamela Harris, Qi Pang, Adrianna Pińska, et al. 2021. *CARTA: Cube Analysis and Rendering Tool for Astronomy.* Astrophysics Source Code Library, record ascl:2103.031, March. ascl: 2103.031.

Crapsi, A., P. Caselli, M. C. Walmsley, and M. Tafalla. 2007. Observing the gas temperature drop in the high-density nucleus of L 1544. 470, no. 1 (July): 221–230. https://doi.org/10.1051/0004-6361:20077613. arXiv: 0705.0471 [astro-ph].

Dirienzo, William J., Crystal Brogan, Rémy Indebetouw, Claire J. Chandler, Rachel K. Friesen, and Kathryn E. Devine. 2015. Physical Conditions of the Earliest Phases of Massive Star Formation: Single-dish and Interferometric Observations of Ammonia and CCS in Infrared Dark Clouds. 150, no. 5 (November): 159. https://doi.org/10.1088/0004-6256/150/5/159. arXiv: 1508.01700 [astro-ph.GA].

Dobashi, Kazuhito, Tomomi Shimoikura, Fumitaka Nakamura, Seiji Kameno, Izumi Mizuno, and Kotomi Taniguchi. 2018. Spectral Tomography for the Line-of-sight Structures of the Taurus Molecular Cloud 1. 864, no. 1 (September): 82. https://doi.org/10.3847/1538-4357/aad62f. arXiv: 1808.01802 [astro-ph.GA].

Dobashi, Kazuhito, Tomomi Shimoikura, Tetsu Ochiai, Fumitaka Nakamura, Seiji Kameno, Izumi Mizuno, and Kotomi Taniguchi. 2019. Discovery of CCS Velocity-coherent Substructures in the Taurus Molecular Cloud 1. 879, no. 2 (July): 88. https://doi.org/10.3847/1538-4357/ab25f0. arXiv: 1905.12773 [astro-ph.GA].

Dunham, Michael M., Stella S. R. Offner, Jaime E. Pineda, Tyler L. Bourke, John J. Tobin, Héctor G. Arce, Xuepeng Chen, et al. 2016. An ALMA Search for Substructure, Fragmentation, and Hidden Protostars in Starless Cores in Chamaeleon I. 823, no. 2 (June): 160. https://doi.org/10.3847/0004-637X/823/2/160. arXiv: 1604.04027 [astro-ph.GA].

Evans, II, Neal J. 1999. Physical Conditions in Regions of Star Formation. 37 (January): 311–362. https://doi.org/10.1146/annurev.astro.37.1.311. arXiv: astro-ph/9905050 [astro-ph].

Friesen, Rachel K., Jaime E. Pineda, co-PIs, Erik Rosolowsky, Felipe Alves, Ana Chacón-Tanarro, Hope How-Huan Chen, et al. 2017. The Green Bank Ammonia Survey: First Results of $NH_3$ Mapping of the Gould Belt. 843, no. 1 (July): 63. https://doi.org/10.3847/1538-4357/aa6d58. arXiv: 1704.06318 [astro-ph.GA].

Fu, Tian-Ming, Yang Gao, and Yu-Qing Lou. 2011. Starless Cloud Core L1517B in Envelope Expansion with Core Collapse. 741, no. 2 (November): 113. https://doi.org/10.1088/0004-637X/741/2/113. arXiv: 1108.2556 [astro-ph.SR].

Galli, P. A. B., L. Loinard, H. Bouy, L. M. Sarro, G. N. Ortiz-León, S. A. Dzib, J. Olivares, et al. 2019. Structure and kinematics of the Taurus star-forming region from Gaia-DR2 and VLBI astrometry. 630 (October): A137. https://doi.org/10.1051/0004-6361/201935928. arXiv: 1909.01118 [astro-ph.SR].

Ginsburg, Adam, and Jordan Mirocha. 2011. *PySpecKit: Python Spectroscopic Toolkit.* Astrophysics Source Code Library, record ascl:1109.001, September. ascl: 1109.001.

Goodman, A. A., P. J. Benson, G. A. Fuller, and P. C. Myers. 1993. Dense Cores in Dark Clouds. VIII. Velocity Gradients. 406 (April): 528. https://doi.org/10.1086/172465.

Henshaw, J. D., A. Ginsburg, T. J. Haworth, S. N. Longmore, J. M. D. Kruijssen, E. A. C. Mills, V. Sokolov, et al. 2019. 'The Brick' is not a brick: a comprehensive study of the structure and dynamics of the central molecular zone cloud G0.253+0.016. 485, no. 2 (May): 2457–2485. https://doi.org/10.1093/mnras/stz471. arXiv: 1902.02793 [astro-ph.GA].

Henshaw, J. D., S. N. Longmore, J. M. D. Kruijssen, B. Davies, J. Bally, A. Barnes, C. Battersby, et al. 2016. Molecular gas kinematics within the central 250 pc of the Milky Way. 457, no. 3 (April): 2675–2702. https://doi.org/10.1093/mnras/stw121. arXiv: 1601.03732 [astro-ph.GA].

Hunter, John D. 2007. Matplotlib: A 2D Graphics Environment. *Computing in Science and Engineering* 9, no. 3 (May): 90–95. https://doi.org/10.1109/MCSE.2007.55.

Kauffmann, J., F. Bertoldi, T. L. Bourke, II Evans N. J., and C. W. Lee. 2008. MAMBO mapping of Spitzer c2d small clouds and cores. 487, no. 3 (September): 993–1017. https://doi.org/10.1051/0004-6361:200809481. arXiv: 0805.4205 [astro-ph].

Kauffmann, Jens, Thushara Pillai, and Paul F. Goldsmith. 2013. Low Virial Parameters in Molecular Clouds: Implications for High-mass Star Formation and Magnetic Fields. 779, no. 2 (December): 185. https://doi.org/10.1088/0004-637X/779/2/185. arXiv: 1308.5679 [astro-ph.GA].

Kirk, J. M., D. Ward-Thompson, and R. M. Crutcher. 2006. SCUBA polarization observations of the magnetic fields in the pre-stellar cores L1498 and L1517B. 369, no. 3 (July): 1445–1450. https://doi.org/10.1111/j.1365-2966.2006.10392.x. arXiv: astro-ph/0603785 [astro-ph].

Koley, Atanu. 2022. Studying the chemical and kinematical structures of dense cores TMC-1C, L1544, and TMC-1 in the Taurus molecular cloud using CCS and NH$_3$ observations. 516, no. 1 (October): 185–196. https://doi.org/10.1093/mnras/stac1935. arXiv: 2208.00968 [astro-ph.GA].

———. 2023. Turbulence measurements in the neutral ISM from HI-21 cm emission-absorption spectra. 40 (September): e046. https://doi.org/10.1017/pasa.2023.43. arXiv: 2308.01808 [astro-ph.GA].

Koley, Atanu, and Nirupam Roy. 2019. Estimating the kinetic temperature from H I 21-cm absorption studies: correction for turbulence broadening. 483, no. 1 (February): 593–598. https://doi.org/10.1093/mnras/sty3152. arXiv: 1811.07352 [astro-ph.GA].

Koley, Atanu, Nirupam Roy, Karl M. Menten, Arshia M. Jacob, Thushara G. S. Pillai, and Michael R. Rugel. 2021. The magnetic field in the dense photodissociation region of DR 21. 501, no. 4 (March): 4825–4836. https://doi.org/10.1093/mnras/staa3898. arXiv: 2012.08253 [astro-ph.GA].

Koley, Atanu, Nirupam Roy, Emmanuel Momjian, Anuj P. Sarma, and Abhirup Datta. 2022. Magnetic field measurement in TMC-1C using 22.3 GHz CCS Zeeman splitting. 516, no. 1 (October): L48–L52. https://doi.org/10.1093/mnrasl/slac085. arXiv: 2207.12604 [astro-ph.GA].

Krieger, Nico, Jürgen Ott, Henrik Beuther, Fabian Walter, J. M. Diederik Kruijssen, David S. Meier, Elisabeth A. C. Mills, et al. 2017. The Survey of Water and Ammonia in the Galactic Center (SWAG): Molecular Cloud Evolution in the Central Molecular Zone. 850, no. 1 (November): 77. https://doi.org/10.3847/1538-4357/aa951c. arXiv: 1710.06902 [astro-ph.GA].

Kudoh, Takahiro, and Shantanu Basu. 2014. Induced Core Formation Time in Subcritical Magnetic Clouds by Large-scale Trans-Alfvénic Flows. 794, no. 2 (October): 127. https://doi.org/10.1088/0004-637X/794/2/127. arXiv: 1408.5648 [astro-ph.SR].

Mangum, Jeffrey G., and Yancy L. Shirley. 2015. How to Calculate Molecular Column Density. 127, no. 949 (March): 266. https://doi.org/10.1086/680323. arXiv: 1501.01703 [astro-ph.IM].

Markwardt, C. B. 2009. Non-linear Least-squares Fitting in IDL with MPFIT. In *Astronomical data analysis software and systems xviii*, edited by D. A. Bohlender, D. Durand, and P. Dowler, 411:251. Astronomical Society of the Pacific Conference Series. September. https://doi.org/10.48550/arXiv.0902.2850. arXiv: 0902.2850 [astro-ph.IM].

Megías, A., I. Jiménez-Serra, J. Martín-Pintado, A. I. Vasyunin, S. Spezzano, P. Caselli, G. Cosentino, and S. Viti. 2023. The complex organic molecular content in the L1517B starless core. 519, no. 2 (February): 1601–1617. https://doi.org/10.1093/mnras/stac3449. arXiv: 2211.16119 [astro-ph.GA].

Myers, P. C., D. Mardones, M. Tafalla, J. P. Williams, and D. J. Wilner. 1996. A Simple Model of Spectral-Line Profiles from Contracting Clouds. 465 (July): L133. https://doi.org/10.1086/310146.

Pillai, T., F. Wyrowski, S. J. Carey, and K. M. Menten. 2006. Ammonia in infrared dark clouds. 450, no. 2 (May): 569–583. https://doi.org/10.1051/0004-6361:20054128. arXiv: astro-ph/0601078 [astro-ph].

Pineda, Jaime E., Jorma Harju, Paola Caselli, Olli Sipilä, Mika Juvela, Charlotte Vastel, Erik Rosolowsky, et al. 2022. An Interferometric View of H-MM1. I. Direct Observation of NH$_3$ Depletion. 163, no. 6 (June): 294. https://doi.org/10.3847/1538-3881/ac6be7. arXiv: 2205.01201 [astro-ph.GA].

Pineda, Jaime E., Anika Schmiedeke, Paola Caselli, Steven W. Stahler, David T. Frayer, Sarah E. Church, and Andrew I. Harris. 2021. Neutral versus Ion Line Widths in Barnard 5: Evidence for Penetration by Magnetohydrodynamic Waves. 912, no. 1 (May): 7. https://doi.org/10.3847/1538-4357/abebdd. arXiv: 2104.12411 [astro-ph.GA].

Robitaille, Thomas, and Eli Bressert. 2012. *APLpy: Astronomical Plotting Library in Python.* Astrophysics Source Code Library, record ascl:1208.017, August. ascl: 1208.017.

Roy, Nirupam, Abhirup Datta, Emmanuel Momjian, and Anuj P. Sarma. 2011. Imaging of the CCS 22.3 GHz Emission in the Taurus Molecular Cloud Complex. 739, no. 1 (September): L4. https://doi.org/10.1088/2041-8205/739/1/L4. arXiv: 1106.4011 [astro-ph.GA].

Ruoskanen, J., J. Harju, M. Juvela, O. Miettinen, A. Liljeström, M. Väisälä, T. Lunttila, and S. Kontinen. 2011. Mapping the prestellar core Ophiuchus D (L1696A) in ammonia. 534 (October): A122. https://doi.org/10.1051/0004-6361/201117862. arXiv: 1109.5070 [astro-ph.GA].

Sahu, Dipen, Sheng-Yuan Liu, Tie Liu, II Evans Neal J., Naomi Hirano, Ken'ichi Tatematsu, Chin-Fei Lee, et al. 2021. ALMA Survey of Orion Planck Galactic Cold Clumps (ALMASOP): Detection of Extremely High-density Compact Structure of Prestellar Cores and Multiple Substructures Within. 907, no. 1 (January): L15. https://doi.org/10.3847/2041-8213/abd3aa. arXiv: 2012.08737 [astro-ph.GA].

Schnee, S., P. Caselli, A. Goodman, H. G. Arce, J. Ballesteros-Paredes, and K. Kuchibhotla. 2007. TMC-1C: An Accreting Starless Core. 671, no. 2 (December): 1839–1857. https://doi.org/10.1086/521577. arXiv: 0706.4115 [astro-ph].

Schnee, S., and A. Goodman. 2005. Density and Temperature Structure of TMC-1C from 450 and 850 Micron Maps. 624, no. 1 (May): 254–266. https://doi.org/10.1086/429156. arXiv: astro-ph/0502024 [astro-ph].

Schnee, Scott, Melissa Enoch, Alberto Noriega-Crespo, Jack Sayers, Susan Terebey, Paola Caselli, Jonathan Foster, et al. 2010. The Dust Emissivity Spectral Index in the Starless Core TMC-1C. 708, no. 1 (January): 127–136. https://doi.org/10.1088/0004-637X/708/1/127. arXiv: 0911.0892 [astro-ph.GA].

Sepúlveda, Inma, Robert Estalella, Guillem Anglada, Rosario López, Angels Riera, Gemma Busquet, Aina Palau, José M. Torrelles, and Luis F. Rodríguez. 2020. VLA ammonia observations of L1287. Analysis of the Guitar core and two filaments. 644 (December): A128. https://doi.org/10.1051/0004-6361/202037895. arXiv: 2011.01651 [astro-ph.GA].

Sipilä, O., P. Caselli, E. Redaelli, and S. Spezzano. 2022. Chemistry and dynamics of the prestellar core L1544. 668 (December): A131. https://doi.org/10.1051/0004-6361/202243935. arXiv: 2209.14025 [astro-ph.GA].

Stahler, Steven W., and Francesco Palla. 2004. *The Formation of Stars.*

Tafalla, M., P. C. Myers, P. Caselli, and C. M. Walmsley. 2004a. On the internal structure of starless cores. I. Physical conditions and the distribution of CO, CS, N$_2$H$^+$, and NH$_3$ in L1498 and L1517B. 416 (March): 191–212. https://doi.org/10.1051/0004-6361:20031704.

———. 2004b. On The Internal Structure Of Starless Cores. Physical and Chemical Properties of L1498 and L1517B. 292, no. 1 (August): 347–354. https://doi.org/10.1023/B:ASTR.0000045036.76044.bd. arXiv: astro-ph/0401148 [astro-ph].

Tokuda, Kazuki, Kakeru Fujishiro, Kengo Tachihara, Tatsuyuki Takashima, Yasuo Fukui, Sarolta Zahorecz, Kazuya Saigo, et al. 2020. FRagmentation and Evolution of Dense Cores Judged by ALMA (FREJA). I. Overview: Inner ∼1000 au Structures of Prestellar/Protostellar Cores in Taurus. 899, no. 1 (August): 10. https://doi.org/10.3847/1538-4357/ab9ca7. arXiv: 2006.06361 [astro-ph.GA].

Walmsley, C. M., and H. Ungerechts. 1983. Ammonia as a molecular cloud thermometer. 122 (June): 164–170.

**Appendix 1.　Calculation of velocity gradient:**
We calculate the local velocity gradient in each pixel as well as the overall velocity gradient in this region based on the method described below. Firstly, we calculate the local velocity gradient of each pixel using its neighbouring pixels according to their weightages, which are based on the distance from that particular pixel and the uncertainties of their center velocities. For example, while calculating the local gradient of a particular pixel *(m,n)*, we take all the other *(i,j)* pixels in this region, and the weight function *f(i,j)* has been given as follows:

$$f_{local}(i,j) = \frac{1}{\Delta V_{c_{i,j}}^2} \cdot \exp\left(-\frac{R_{i,j-m,n}^2}{2 \cdot \left(\frac{4.6}{2.355}\right)^2}\right) \quad (7)$$

Here, $\Delta V_{c_{i,j}}$ is the uncertainty of the center velocity of the *i,j* pixel and $R_{i,j-m,n}$ is the distance between *i,j* and *m,n* pixel. We take the value of the full width at half maximum (FWHM) of the Gaussian function is 4.6 for the following reason. The beam size of this observation is $\sim 4.01''(\theta_{major}) \times 3.37''(\theta_{minor})$ and therefore, the effective beam size is $\sqrt{\theta_{major} \times \theta_{minor}} = 3.68''$. On the other hand, the pixel size in our image cube is $0.80''$. That is why, we take the FWHM is equal to 4.6 pixels.

From the *pyspeckit* result, we obtain the pixel-wise center velocity ($V_c$) and its associated error ($\Delta V_c$). Therefore, the horizontal component of the local velocity gradient of a pixel *(m,n)* is obtained by the formula:

$$\left(\frac{dV_c}{dR}\right)_{m,n\ local,x} = \frac{\sum_{i,j} f_{local}(i,j) \left(\frac{dV_c}{dR}\right)_{i,j-m,n} cos\theta_{i,j-m,n}}{\sum_{i,j} f_{local}(i,j)} \quad (8)$$

Similarly, the vertical component of the local velocity gradient of a pixel *(m,n)* is:

$$\left(\frac{dV_c}{dR}\right)_{m,n\ local,y} = \frac{\sum_{i,j} f_{local}(i,j) \left(\frac{dV_c}{dR}\right)_{i,j-m,n} sin\theta_{i,j-m,n}}{\sum_{i,j} f_{local}(i,j)} \quad (9)$$

Therefore, the magnitude of the local velocity gradient at the pixel *(m,n)* is as follows:

$$\left(\frac{dV_c}{dR}\right)_{m,n\ local} = \sqrt{\left(\frac{dV_c}{dR}\right)_{m,n\ local,x}^2 + \left(\frac{dV_c}{dR}\right)_{m,n\ local,y}^2} \quad (10)$$

and the direction of the local velocity gradient at the pixel *(m,n)* is:

$$\theta_{m,n\ local} = tan^{-1}\left(\frac{\left(\frac{dV_c}{dR}\right)_{m,n\ local,y}}{\left(\frac{dV_c}{dR}\right)_{m,n\ local,x}}\right) \quad (11)$$

On the other hand, for the overall velocity gradient, we first calculate the velocity gradient at each pixel (m,n) using the other pixels (i,j), and then obtain the average value as well as direction of the velocity gradient. In this case, we use the following weight function:

$$f_{overall}(i,j) = \frac{1}{\Delta V_{c_{i,j}}^2} \quad (12)$$

**Appendix 2.　Ratio of rotational energy to gravitational energy for a spherical system of non-uniform density**
Ratio of rotational energy to gravitational energy is defined by the parameter β:

$$\beta = \frac{E_{rot}}{|U|} = \frac{\frac{1}{2}Iw^2}{q\frac{GM^2}{R}} = \frac{1}{2}\frac{p}{q}\frac{w^2R^3}{GM} \quad (13)$$

Here, *I* is the moment of inertia, *w* is the rotational velocity, *G* is the gravitational constant, *R* is the radius of the spherical system, *p* and *q* are unit-less numbers, which vary depending on the density profile of the spherical system (Kauffmann, Pillai, and Goldsmith 2013).

We consider a solid sphere of radius *R*, mass *M*, and density $\rho(r) = K\rho_0 r^{-\alpha}$ (neglect singularity at r=0). Now, for calculating the moment of inertia (*I*), we consider the solid sphere as the sum of consecutive spherical shells of infinitesimal thickness, one of which radius is *r* and thickness is *dr*.

Now, the moment of inertia of that spherical shell is:

$$dI = \left(\frac{8\pi K\rho_0}{3}\right)r^{4-\alpha}dr \quad (14)$$

Then, the total moment of inertia of the solid sphere becomes:

$$I = \left(\frac{2MR^2}{3}\right)\left(\frac{3-\alpha}{5-\alpha}\right) \quad (15)$$

Similarly, for calculating the gravitational potential energy, we consider that a sphere of radius *r* is already formed and we want to add a thin spherical shell on top of that whose thickness is *dr* and mass is *dm(r)*, then potential energy is:

$$dU = -\frac{Gm(r)dm(r)}{r} \quad (16)$$

Now, the total total potential energy becomes:

$$U = -\left(\frac{3-\alpha}{5-2\alpha}\right)\left(\frac{GM^2}{R}\right) \quad (17)$$

For $\alpha = 0$; $I = \frac{2}{5}MR^2$ and $U = -\frac{3}{5}\frac{GM}{R^2}$; so $\frac{p}{q} = 0.66$. Similarly, for $\alpha = 2$; $I = \frac{2}{9}MR^2$ and $U = -\frac{GM}{R^2}$; so $\frac{p}{q} = 0.22$.